\documentclass[aps,prl,groupedaddress,superscriptaddress,longbibliography,
twocolumn,
,amsmath,amssymb]{revtex4-2}
\usepackage{color}
\definecolor{darkblue}{rgb}{0.0, 0.0, 0.55}
\usepackage[colorlinks=true,linkcolor=darkblue,anchorcolor=red,citecolor=darkblue, urlcolor=darkblue]{hyperref}
\usepackage{romannum}
\usepackage{graphicx}
\usepackage{bm}
\usepackage{ulem}
\usepackage{subfigure}
\usepackage[capitalize]{cleveref}
\usepackage{epstopdf}

\begin{document}

\title{Critical Behavior and Universal Signature of an Axion Insulator State}
\author{Hailong Li}
\affiliation{International Center for Quantum Materials, School of Physics,
Peking University, Beijing 100871}
\author{Hua Jiang}
\affiliation{School of Physical Science and Technology, Soochow University, Suzhou 215006, China}
\affiliation{Institute for Advanced Study, Soochow University, Suzhou 215006, China}
\author{Chui-Zhen Chen}
\email{czchen@suda.edu.cn}
\affiliation{School of Physical Science and Technology, Soochow University, Suzhou 215006, China}
\affiliation{Institute for Advanced Study, Soochow University, Suzhou 215006, China}
\author{X. C. Xie}
\email{xcxie@pku.edu.cn}
\affiliation{International Center for Quantum Materials, School of Physics,
Peking University, Beijing 100871}
\affiliation{Beijing Academy of Quantum
Information Sciences, Beijing 100193, China}
\affiliation{CAS Center for
Excellence in Topological Quantum Computation, University of Chinese Academy
of Sciences, Beijing 100190, China}
\date{\today }

\begin{abstract}
Recently, the search for an axion insulator state in the ferromagnet-3D topological insulator (TI) heterostructure and $\mathrm{MnBi_2Te_4}$ has attracted intensive interest. 
However, its detection remains difficult in experiments.
We systematically investigate the disorder-induced phase transition of the axion insulator state in a 3D TI with antiparallel magnetization alignment surfaces. It is found that there exists a 2D disorder-induced phase transition which shares the same universality class with the quantum Hall plateau to plateau transition.
Then, we provide a phenomenological theory which maps the random mass Dirac Hamiltonian of the axion insulator state into the Chalker-Coddington network model.
Therefore, we propose to probe the axion insulator state by investigating the universal signature of such a phase transition in the ferromagnet-3D TI heterostructure and $\mathrm{MnBi_2Te_4}$.
Our findings not only show a global phase diagram of the axion insulator state, but also provide a new experimental routine to probe it. 
\end{abstract}


  \maketitle

{\textit{Introduction.}}---Topology and symmetry breaking play a key role in describing phases of matter. 
Since proposed in 2008, the axion insulator state has attracted extensive experimental and theoretical studies~\cite{rmp_zhang,tft_prb,TME_nagaosa,analytic_TME,nagaosa_disorderAI,1FMTI_prb,2FMTI_prl,axion_experiment2017}. The 3D topological insulator (TI) is a time-reversal symmetry-protected topological matter characterized by gapless Dirac surface states in the bulk gap~\cite{rmp_zhang}. If the time-reversal symmetry is broken, an axion insulator state shows up when the gapless Dirac surface states are gapped out by the magnetizations pointing outwards (inwards) the surfaces~\cite{rmp_zhang,tft_prb}. In comparison with a trivial insulator state, the axion insulator state possesses unique electromagnetic response from the massive Dirac surface states, giving rise to novel phenomena such as quantized topological magnetoelectric effect (TME) and half-quantized surface Hall conductance~\cite{tft_prb,TME_nagaosa,analytic_TME,nagaosa_disorderAI}. In the experiment, the axion insulator shows huge longitudinal resistance and zero Hall conductance, because the top and bottom surface Hall conductance cancels out~\cite{wangyayu,2FMTI_prl,axion_experiment2017}. These results, however, are coincident with a trivial band insulator. Therefore, a definitive experimental evidence for the axion insulator state is still missing.

\begin{figure}[htbp]
  \includegraphics[width=\columnwidth]{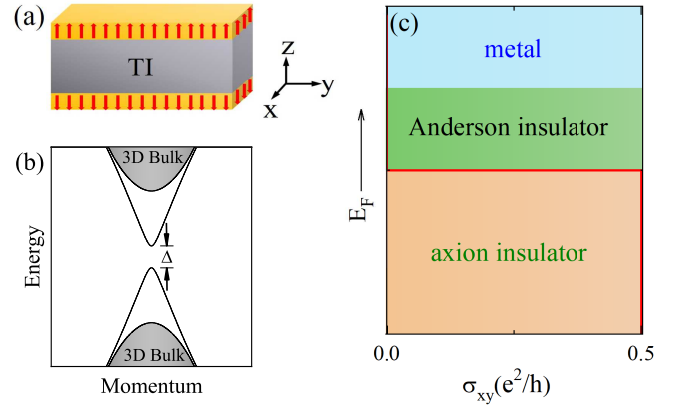}
  \caption{(color online) (a) Schematic plot of an axion insulator consisting of a 3D time-reversal invariant TI with antiparallel magnetization alignment surfaces. (b) The magnetization term can open a gap $\Delta$ at the Dirac point of the surface states. (c) Schematic phase diagram of the 3D TI with antiparallel magnetization alignment surfaces under weak disorder. The red curve depicts the Hall conductance varying with the Fermi energy.
  }
  \label{Fig1}
\end{figure}
On the other side, the characteristic properties of Anderson phase transitions in disordered systems, especially the critical exponents, depend only on general properties of the model, such as spatial dimensionality, symmetry, etc. The experimental studies of the quantum phase transition have already been extensively performed for magnetic TIs and revealed unique properties of topological states ~\cite{Checkelsky2014,kou2015,Wang2015,Chang2016,qahtoAI,wangjian,zhangyuanbo,wangyayu}. Hence, in this work, we propose to probe the axion insulator state by shedding light on the disorder-induced metal-insulator transition in 3D magnetic TIs.
In particular, we systematically study the disorder effect of a 3D TI with antiparallel magnetization alignment surfaces and give a global phase diagram of the axion insulator. Such a surface magnetization has been realized in the ferromagnet-3D TI heterostructure and $\mathrm{MnBi_2Te_4}$.
Notably, there exists a 2D quantum-Hall-type phase transition between the axion insulating phase and the Anderson insulating phase, which provides a universal experimental signature of an axion insulator state [see \cref{Fig1}(c)]. To be specific, with weak disorder and increase in the Fermi energy, the axion insulator will undergo a 2D delocalized transition, become an Anderson insulator and then transform into a diffusive metal after a 3D insulator-metal transition.
We also provide a phenomenological theory that relates the disordered axion insulator to the Chalker-Coddington network model. Furthermore, the 2D phase transition remains in the presence of bulk antiferromagnetism for $\mathrm{MnBi_2Te_4}$, and thereby it is model-independent.
The universal phase transition behavior of the axion insulator we predicted can be detected in the ferromagnet-3D TI heterostructure as well as the antiferromagnetic TI $\mathrm{MnBi_2Te_4}$~\cite{mnbite_cpl,mnbite_qahtransition,mnbitefamily,AFMTI_nature,wangyayu,wangjian,zhangyuanbo,dassarma,mnbite_prx}.

{\textit{Effective model of the axion insulators.}}---To start with, we consider a 3D TI with antiparallel magnetization alignment surfaces [see Fig.\ref{Fig1}(a)] that has been realized in experiments ~\cite{1FMTI_prb,2FMTI_prl,wangyayu} and the four-band effective Hamiltonian can be written as
\begin{equation}
H = H_{0} + H_M
\end{equation}
where $H_{0}(\mathbf{k})=\sum_{i=1}^{4} d_{i}(\mathbf{k}) \Gamma_{i}$ with $d_{1}=A_{1}k_{x}$, $d_{2}=A_{1}k_{y}$, $d_{3}=A_{2}k_{z}$, and $d_{4}=M_0-B_{1} k_{z}^{2}-B_{2}\left(k_{x}^{2}+k_{y}^{2}\right)$. It describes a time-reversal invariant TI around $\Gamma$ point and hosts a single Dirac cone on each surface~\cite{rmp_zhang}.
Here, $A_i$ and $B_i$ are the model parameters and $M_0$ controlls the bulk gap of the 3D TI.
$\Gamma_{i}=s_{i} \otimes \sigma_{1}$ for $i=1,2,3$, and $\Gamma_{4}=s_{0} \otimes \sigma_{3}$. $s_{i}$ and $\sigma_{i}$ are the Pauli matrices for the spin and orbital degrees of freedom.
The Zeeman splitting $H_M=M(z)s_{z} \otimes \sigma_{0}$ where $M(z)$ takes the values $\pm M_{z}$ on the top and bottom surfaces, respectively, and zero elsewhere [see Fig.~\ref{Fig1}(a)].
Such a time-reversal breaking mass term will open up a Dirac gap $\Delta \approx 2|M_z|$ on the top/bottom surface [see Fig.~\ref{Fig1}(b)], which is described by $H_{surf}^{t/p}=A_1(\sigma_x k_x + \sigma_x k_x) \pm M_z\sigma_z$ and
leads to a half quantized Hall conductance $\sigma_{xy}^{t/b}=\pm\frac{e^{2}}{2 h}$ for the top/bottom surfaces~\cite{tft_prb}.
Note that we have demonstrated the existence of an axion insulating phase by numerically calculating half-quantized surface Hall conductance and discretizing Hamiltonian $H$ on square lattices [see Sec. S1 of \cite{SM}].

Unlike the quantum anomalous Hall insulator, the total Hall conductance of the axion insulator is zero, i.e. $\sigma_{xy}^{t}+\sigma_{xy}^{b}=0$ and there are no chiral edge modes within the massive Dirac surface gap. This explains why a zero Hall conductance plateau and a huge longitudinal resistance were observed for the axion insultor state in the recent experiments~\cite{wangyayu,2FMTI_prl,axion_experiment2017}.
However, these results are coincident with a trivial band insulator.
To unveil signatures of the axion insulator, we will investigate the disorder-induced critical behavior of the axion insulator in the following. We include the random magnetic disorder as $H_D=V(\mathbf{r})s_{z} \otimes \sigma_{0}$
where $V(\mathbf{r})$ is uniformly distributed within $[-W/2,W/2]$ with $W$ denoting the disorder strength.


\begin{figure}[b]
  \centering
  \includegraphics[width=\columnwidth]{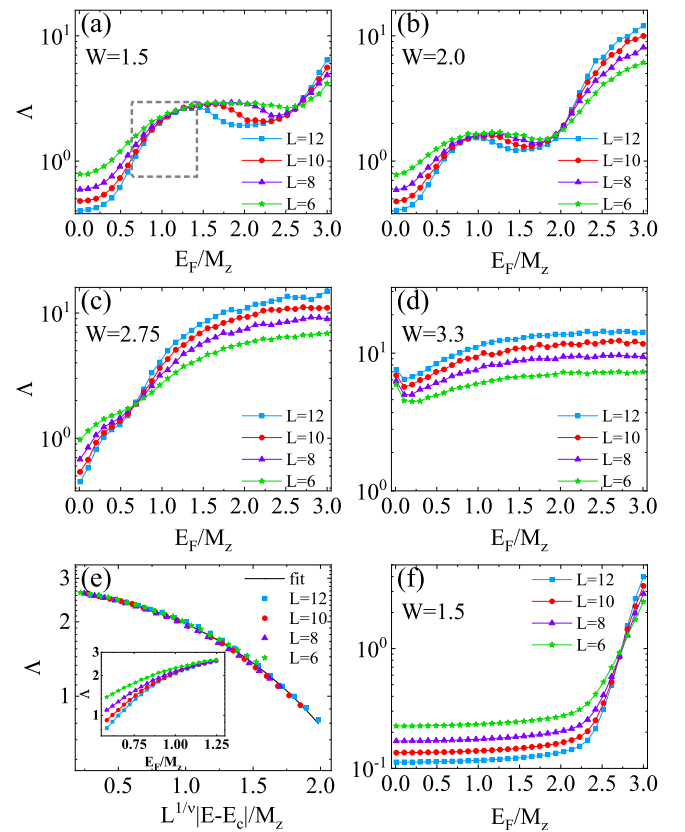}
  \caption{(color online) (a)-(d) Renormalized localization length $\Lambda=\lambda(L)/L$ against the Fermi energy $E_{F}$ at different magnetic disorder strengths $W$. The curves correspond to different sample widths $L$. Here, the $x$ direction is periodic and the $z$ direction is open. (e) It shows a fit of the numerical data in the inset with $W=1.5$ by a scaling function $\Lambda=f(L^{1/\nu}|E-E_c|/M_{z})$. The polynomial fitting method gives a critical exponent $\nu=2.654\pm 0.213$ and a critical Fermi energy $E_c=0.164\pm 0.005$. Here, the raw data is obtained from the grey rectangle region of (a). (f) $\Lambda$ with periodic boundary condictions in the $x$ and $z$ directions and $W=2.0$. Other parameters are fixed as $A_{1}=A_{2}=0.55$, $B_{1}=B_{2}=0.25$, $M_0=0.3$, and $M_{z}=0.12$.}
  \label{Fig2}
\end{figure}

{\textit{Quantum-Hall-type phase transition.}}---
To calculate the localization length, we consider a 3D long bar sample of length $L_y$ and widths $L_{x}=L_{z}=L$ with periodic boundary condition in the $x$ direction and open boundary condition in the $z$ direction unless otherwise specified. The localization length $\lambda(L)$ is obtained by using the transfer matrix method~\cite{Localization1,Localization2,Localization3}. Generally, criticality can be accessed from the renormalized localization length $\Lambda=\lambda/L$, which increases with $L$ in a metallic phase, decreases with $L$ in an insulating phase, and does not depend on $L$ at the critical point of a phase transition.

We consider a sample like Fig.~\ref{Fig1}(a) and perform a finite-size scaling analysis as shown in Fig.~\ref{Fig2}. It shows that the axion insulator undergoes multiple phase transitions with increasing Fermi energy $E_F$ under different disorder strengths $W$. To be specific, for weak disorder $W=1.5$ and $2$ in Figs.~\ref{Fig2}(a)-(b), one can identify an axion insulator phase with $d\Lambda/dL<0$ when $|E_F/M_z|\lesssim 1$, where the Fermi energy $E_F$ is within the the surface Dirac gap $M_z$.
With increasing Fermi energy,
we find that the system goes delocalized with $d\Lambda/dL=0$ and arrive at an Anderson insulator phase without half-quantized Hall conductance as we will show below. This is reminiscent of the plateau-plateau transition in 2D quantum Hall system.
To identify that such a phase transition belongs to the quantum-Hall type, we perform a single-parameter scaling analysis by using a polynomial fitting method in \cref{Fig2}(e)~\cite{scaling_function}. The critical exponent $\nu$ which is expected to exhibit universality is extracted as $\nu=2.654\pm 0.213$ and a critical Fermi energy $E_c=0.164\pm 0.005$. This estimate of $\nu$ is in agreement with recent numerical results, $\nu\sim 2.6$, based on the Chalker-Coddington model of the integer quantum Hall effect~\cite{qh1,qh2,qh3}.
Moreover, to verify that the phase transition comes from two-dimensional Dirac surfaces, we take periodic boundary conditions in the both $x$ and $z$ directions, and calculate $\Lambda$ in \cref{Fig2}(f).
In comparison to \cref{Fig2}(a), one can see the lower 2D delocalized state disappears and only the higher critical point remains, which indicates a 3D Anderson metal-insulator transition.
As a result, we conclude that there exists a universal 2D quantum-Hall-type phase transition in the axion insulator which thus provide a universal signature of the axion insulator state in experiment [see Sec. S2 of~\cite{SM}].
Finally, in the large disorder limit, the axion insulator is gradually suppressed by the 3D critical point, and eventually disappears [see Figs.~\ref{Fig2}(a)-(d)] and the system becomes a 3D diffusive metal. This can be seen more easily in the phase diagram in the following [see Fig. 3(d)].


\begin{figure}[t]
  \centering
  \includegraphics[width=\columnwidth]{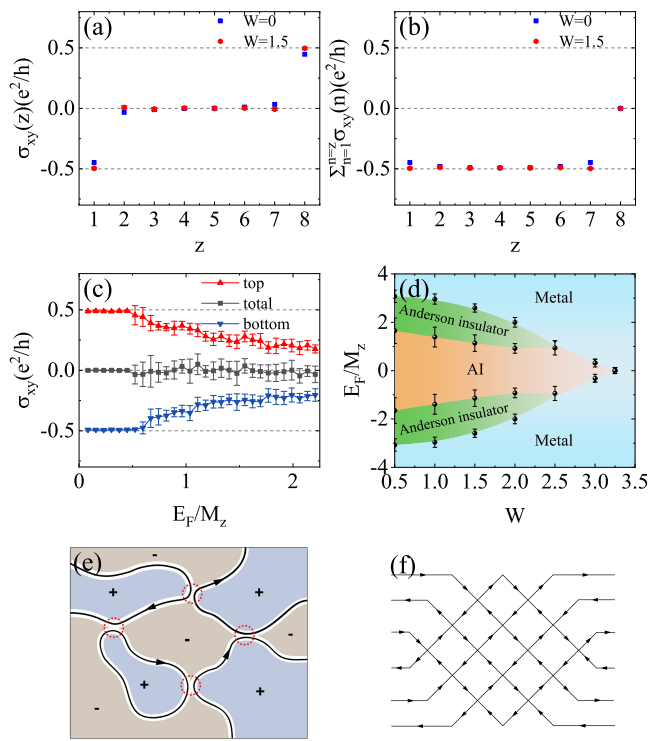}
  \caption{(color online) (a) It shows the Hall conductance as a function of the layer index $z$ with $z=1$ for the bottom layer and $z=8$ for the top layer. (b) The cumulative summation of the Hall conductance from $z=1$ (bottom surface) to the $z$-th layer is disorder-averaged. (c) The Hall conductance of the top surface states ($\sum_{n=7}^{n=8}\sigma_{xy}(n)$), the bottom surface states ($\sum_{n=1}^{n=2}\sigma_{xy}(n)$) and the whole sample ($\sum_{n=1}^{n=8}\sigma_{xy}(n)$) is shown by varying the Fermi energy. (d) Phase diagram of a disordered axion insulator (AI) in~\cref{Fig1}(a) in the $E_{F}/M_{z}-W$ plane. (e) Chiral edge states along domain walls of positive ($+$) and negative ($-$) masses of Dirac fermions. (f) The Chalker-Coddington network model on a quasi-1D system, where a scattering matrix is used to describe the scattering from two incoming to two outgoing modes at each node (crossing point). }
  \label{Fig3}
\end{figure}

{\textit{Hall conductance and phase diagram.}}---
Next, we investigate the Hall conductance of the system to further identify the two insulating phases illustrated above.
For a low Fermi energy and if the system is in an axion insulating state, the surface Hall conductance and the net Hall conductance of the whole sample are expected to be half-quantized and zero respectively. On the other hand, when the Fermi energy goes across the 2D delocalized state, the system is converted into an Anderson insulator and the surface Hall conductance is expected to lose its half-integer quantization and approach zero.

Here, the layer-dependent Hall conductance is evaluated by a noncommutative real-space Kubo formula~\cite{chern1,chern2},
\begin{equation}
  \sigma_{xy}(z)=\frac{2\pi i e^2}{h} \left\langle\operatorname{Tr}\left\{P\left[-i\left[\hat{x}, P\right],-i\left[\hat{y}, P\right]\right]\right\}_z\right\rangle_{W} \label{eq:chern}
\end{equation}
with periodic boundary conditions in both $x$ and $y$ directions. $\left\langle...\right\rangle_{W}$ represents the disorder average, $(\hat{x},\hat{y})$ denotes the position operator and $\operatorname{Tr}\left\{...\right\}_z$ is trace over the $z$th layer. $P$ is the projector onto the occupied states of $H$. Through \cref{eq:chern}, we calculate the layer-dependent Hall conductance $\sigma_{xy}(z)$ as a function of the layer index $z$ [see \cref{Fig3}(a)]. For both a clean sample ($W=0$) and a disordered sample ($W=1.5$), the non-zero Hall conductance mainly comes from surfaces near the top ($z=1$) and bottom ($z=8$) layers,
while it decays exponentially into bulk (between $z=2$ and $z=7$).
This is coincident with the exponential decay of the surface states in 3D TI, where the surface states exist in several layers near the surfaces~\cite{surface_state_3DTI}.
To gain further insight into the Hall conductance of the system, we take the total Hall conductance of the several layers into consideration.
The cumulative summation of the contribution from each layer, i.e., $\sum_{n=1}^{n=z}\sigma_{xy}(n)$, is shown in \cref{Fig3}(b). For $W=1.5$ ($W=0$), the cumulative Hall conductance becomes half-quantized at $z=1$ ($z=2$) indicating that the bottom surface Hall conductance is half-quantized. Besides, when $z=8$, the net Hall conductance of the whole sample vanishes for both cases since the half-quantized Hall conductance of the top and bottom surfaces cancel each other out.
Moreover, in Fig~3.(c), one can see a half-integer quantized surface Hall conductance plateau
for the axion insulator at the lower energy regime ($|E_F/M_z|\lesssim 0.5$) while it gradually approaches zero for the Anderson insulator at the high energy regime as expected.

Now, we perform finite-size scaling analysis under various disorder strengths and summarize a phase diagram in the $E_{F}/M_{z}-W$ plane [see Fig.~3(d)]. Due to the particle-hole symmetry of the Hamiltonian $H$, the phase diagram is symmetric about $E_{F}/M_{z}=0$.
For weak $W$, the axion insulator and the 3D diffusive metal phases are separated by the Anderson insulator phase.
 With increase of the disorder strength, they gradually draw close to each other and are connected eventually at a large disorder strength $W\approx 2.5$. Then, the bulk gap continues to shrink and close at about $W\approx 3.2$, and the sample ends up as a 3D metal, which manifests the features of levitation and pair annihilation~\cite{levitation}.


To understand the underlying mechanism for the above 2D phase transition, we provide a phenomenological explanation. When $W=0$, the surface state satisfies a 2D Dirac Hamiltonian with homogenous mass in the real space. With increase of the random magnetic (mass) disorder $W$, the surface mass term becomes spatially inhomogenous such that the Dirac fermions with positive and negative masses coexist. As shown in \cref{Fig3}(e), there exists a chiral edge state between two regions with Dirac masses of opposite signs~\cite{ludwig}. Moreover, the 2D random Dirac mass Hamiltonian can be mapped onto the Chalker-Coddinton model which can describe the quantum-Hall plateau to plateau transition~\cite{ludwig,chalker1,chalker2}. \cref{Fig3}(f) shows the Chalker-Coddington network model on a quasi-1D system. At each node, the incoming and outgoing channels denotes the chiral edge modes confined at domain walls between positive and negative masses of Dirac fermions [as shown in \cref{Fig3}(e)]. Thus, the Chalker-Coddington model is equivalent to the random Dirac Hamiltonian~\cite{ludwig,chalker1,chalker2}. Morever, the critical exponent $\nu=2.654\pm 0.213$ in our model is in good agreement with previous numerical results $\nu\sim 2.6$, based on the Chalker-Coddington model ~\cite{qh1,qh2,qh3}. Consequently, the aforementioned quantum phase transition from the axion insulator to the Anderson insulator shares the same universality class with the quantum Hall plateau to plateau transition.

{\textit{Discussions and experimental routines.}}---
Recently, the axion insulator state characterized by a zero Hall conductance plateau and a huge longitudinal resistance was reported experimentally in the ferromagnet-3D TI heterostructure and the antiferromagnetic TI $\mathrm{MnBi_2Te_4}$ ~\cite{qahtoAI,wangyayu,2FMTI_prl,axion_experiment2017}.
Furthermore, they found a phase transition from the axion insulator state to the Chern insulator state as the Hall conductance increased from zero to $e^2/h$,  when the magnetizations of the top and bottom surfaces were driven from an antiparallel alignment to a parallel alignment by sweeping an external magnetic field ~\cite{qahtoAI,wangyayu}.
However, these results are also coincident with the trivial band insulator that was  previously  reported in magnetic TI system \cite{kou2015,Wang2015}.
In Fig.~4(a), if we start with the axion insulator phase for antiparallel alignment configuration ($M_zM_b<0$), both the two-terminal conductance ~\cite{numerical01,numerical02,numerical03} and the Hall conductance increase from zero to $e^2/h$ by the magnetic flipping of the bottom surface. This demonstrates the axion insulator to Chern insulator transition as observed in the experiments. %
On the contrary, for the Anderson insulator, the two-terminal conductance keeps small and the Hall conductance does not show any quantized behavior [see Fig.~4(b)].
Thus, we conclude that these reported experimental systems are probably in the axion insulating phase but cannot be in the Anderson insulating phase. They are good candidate materials to further identify the axion insulator by probing the universal 2D phase transition, thereby ruling out the trivial band insulator.

Regarding the difficulty to vary the Fermi energy in a 3D sample, we suggest applying an in-plane magnetic field in the $x$ ($y$) direction. It will increase $E_F/M_z$ by reducing the out-of-plane magnetization $M_{z}$, and the 2D phase transition can appear~\footnote{Although an in-plane magnetic field can not only reduce the out-of-plane component of the magnetization but also increase the in-plane component of it, the surface gap is only determined by the out-of-plane component of the magnetization and is unaffected the in-plane component.
Therefore, the decrease in $M_z$ is equivalent to the increase in $E_F/M_z$.}.
Moreover, we further include the bulk antiferromagnetism of the Hamiltonian $H$ in Eq.(1) as  the effective model of $\mathrm{MnBi_2Te_4}$ and repeat the finite-size scaling as shown in Fig. S4 of the supplementary material~\cite{dassarma,MnBiTe_FPC_wangjing,luhaizhou_nolocal}. The 2D phase transition remains and thereby is model-independent [see Sec. S3 of~\cite{SM}].
Therefore, we propose to probe the universal 2D phase transition of the axion insulator in a ferromagnet-3D TI heterostructure or an antiferromagnetic TI $\mathrm{MnBi_2Te_4}$.

\begin{figure}[t]
  \centering
  \includegraphics[width=\columnwidth]{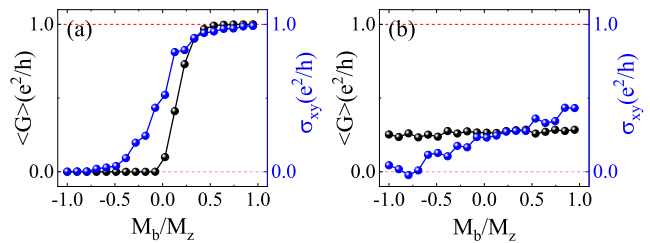}
  \caption{(color online) Disorder-averaged conductance $\langle G \rangle $ and total Hall conductance $\sigma_{xy}$ as functions of the magnetization in the bottom surface $M_b$. (a) The axion insulating phase with $E_{F}/M_{z}\approx0.083$. (b) The Anderson insulating phase with $E_{F}/M_{z}=2$. Other parameters: the disorder strength $W=1.5$ and the system size $L_{z}\times L_{x}\times L_{y}=8\times 40 \times 400$.}
  \label{Fig4}
\end{figure}

{\textit{Conclusion.}}---
To summarize, we investigate the disorder-induced Anderson transition of an axion insulator and find a 2D phase transition between the axion insulating phase and the Anderson insulating phase, which does not occur in trivial band insulators. The 2D phase transition originates from the 2D massive Dirac Hamiltonian which lives in the surface of a 3D system.
From the viewpoint of Chalker-Coddington network model, an exponent $\nu \approx 2.65$ via the finite-size scaling analysis strongly suggests the 2D phase transition from the axion insulator to the Anderson insulator shares the same universality class with the quantum Hall plateau to plateau transition. Therefore, we propose to probe the axion insulator state
by investigating the universal signature of 2D quantum-Hall-type critical behaviors in 3D magnetic TIs.

{\textit{Acknowledgement.}}---
This work is financially supported by NSFC (Grants Nos. 11534001, 110801719 and 11822407), the Strategic Priority Research Program of Chinese Academy of Sciences (Grant No. XDB28000000) and the Priority Academic Program Development of Jiangsu Higher Education Institutions.
C.-Z.C. are also funded by NSFC of Jiangsu province BK20190813.

{\textit{Note added.}---
Toward the completion of this work, we became aware of an independent study~\cite{song2020delocalization} which focuses on similar topics but different aspects. 

\end{document}


\title{Supplementary Materials for ``Critical Behavior and Universal Signature of an Axion Insulator State''}
\author{Hailong Li}
\affiliation{International Center for Quantum Materials, School of Physics,
Peking University, Beijing 100871}
\author{Hua Jiang}
\affiliation{School of Physical Science and Technology, Soochow University, Suzhou 215006, China}
\affiliation{Institute for Advanced Study, Soochow University, Suzhou 215006, China}
\author{Chui-Zhen Chen}
\email{czchen@suda.edu.cn}
\affiliation{School of Physical Science and Technology, Soochow University, Suzhou 215006, China}
\affiliation{Institute for Advanced Study, Soochow University, Suzhou 215006, China}
\author{X. C. Xie}
\email{xcxie@pku.edu.cn}
\affiliation{International Center for Quantum Materials, School of Physics,
Peking University, Beijing 100871}
\affiliation{Beijing Academy of Quantum
Information Sciences, Beijing 100193, China}
\affiliation{CAS Center for
Excellence in Topological Quantum Computation, University of Chinese Academy
of Sciences, Beijing 100190, China}
\date{\today }

\maketitle
\tableofcontents
\section{Half-quantized Hall conductance of the 3D TI with antiparallel magnetization alignment surfaces in Fig.~1}
\begin{figure}[htbp]
  \includegraphics[scale=0.3]{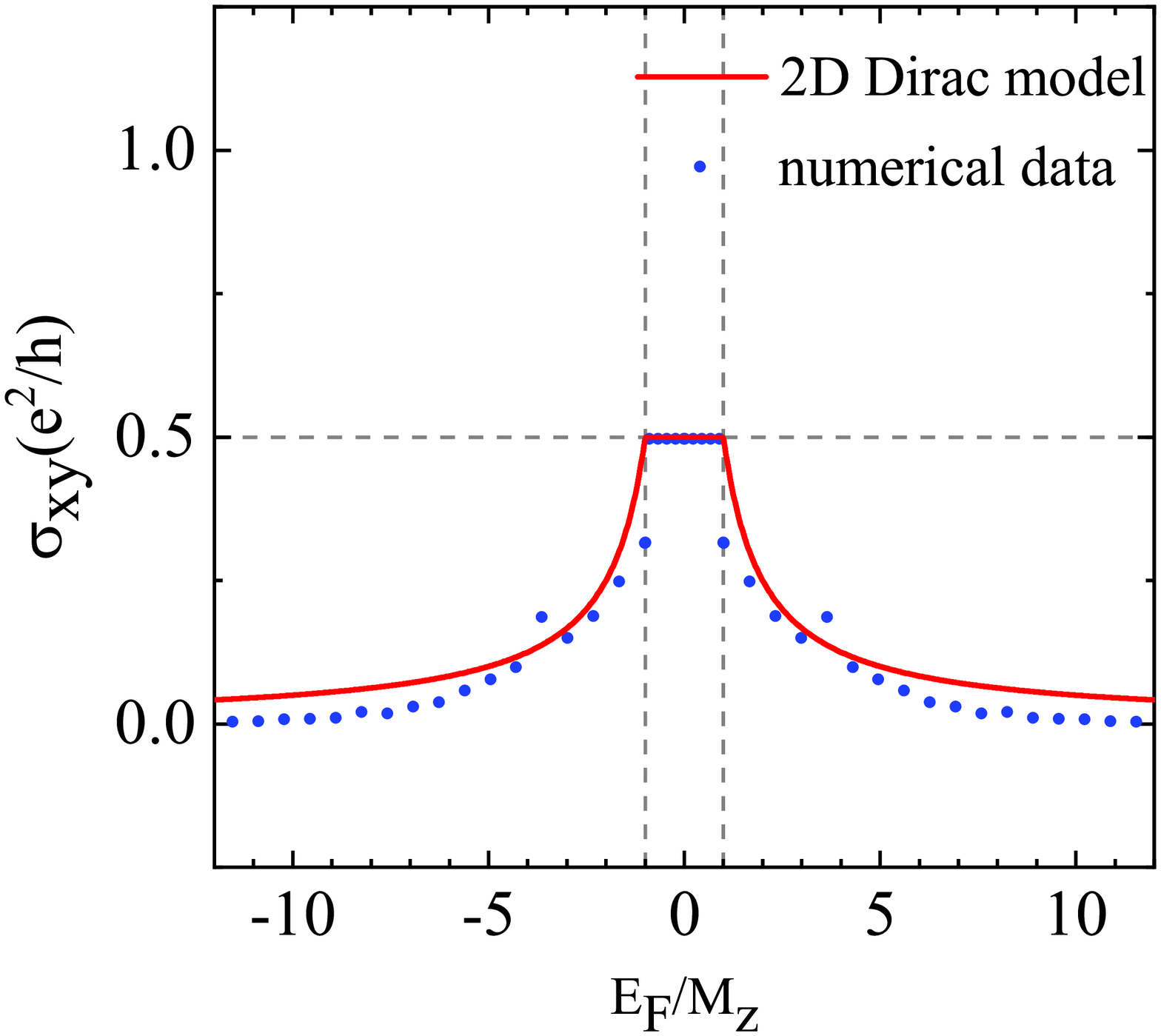}
  \caption{It shows numerical simulations of the Hall conductance of the top surface. The red curve is obtained from a 2D massive Dirac Hamiltonian. Here, the system size is $L_x\times L_y \times L_z = 30\times30\times8$, $E_F$ is the Fermi energy and $M_z$ the magnetization strength. Other parameters are $A_{1}=0.8$, $A_{2}=0.5$, $B_{1}=0.25$, $B_{2}=0.8$, $M_0=0.5$, and $M_{z}=0.15$.
  }
  \label{half_hall_cond}
\end{figure}
Here, we would like to demonstrate the 3D TI with antiparallel magnetization alignment surfaces shown in Fig.~1(a) of the main text describes an axion insulator.
To justify the existence of an axion insulator phase that is featured by the half-quantized-surface Hall conductance, we discretize Hamiltonian $H$ on square lattices and calculate the Hall conductance of the surface states numerically. For a clean system, the numerical result in Fig.~\ref{half_hall_cond} through Eq.~(2) in the main text shows the Hall conductance of the top surface in our model is $e^2/(2h)$ with the Fermi energy $E_F$ in the surface gap, and approaches zero when the Fermi energy $E_F$ moves outside the gap. These results are captured by the analytic curve, which is obtained from the Hall conductance of the 2D massive surface Dirac Hamiltonian $H_{surf}^t=A_1(\sigma_x k_x + \sigma_x k_x) + M_z\sigma_z$, i.e.
\begin{equation}
\sigma_{xy}= \frac{M_{z}}{2|M_{z}|}\Theta(|M_{z}|-|E_{F}|)+\frac{M_{z}}{2|E_{F}|}\Theta(|E_{F}|-|M_{z}|)\nonumber
\end{equation}
with the Heaviside function $\Theta(...)$. Consequently, such a model with an outward pointing magnetization $H$ describes an axion insulator.
\section{Anderson transition in a normal insulator}
\begin{figure}[htbp]
    \includegraphics[width=\columnwidth]{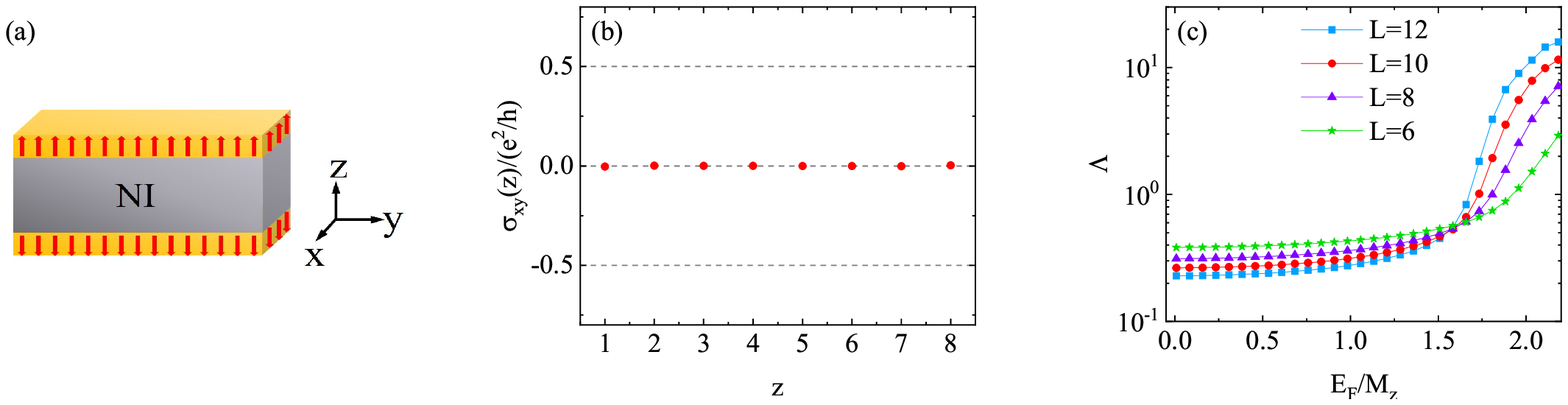}
    \caption{(color online) (a) Schematic plot of a normal insulator with antiparallel magnetization alignment surfaces. (b) It shows the Hall conductance as a function of the layer index $z$ with $z=1$ for the bottom layer and $z=8$ for the top layer. Here, the disorder strength $W=0$ and the Fermi energy $E_F/M_z\approx0.083$. (c) Renormalized localization length $\Lambda=\lambda(L)/L$ against the Fermi energy $E_{F}/M_{z}$ with the disorder strength $W=1.5$. The curves corresponds to different sample widths $L$. Here, the $x$ direction is periodic and the $z$ direction is open.  A new set of parameters to describe a normal insulator is $A_{1}=A_{2}=0.55$, $B_{1}=B_{2}=0.25$, $M_0=-0.3$, and $M_{z}=0.12$.
    }
    \label{Figs1}
  \end{figure}
As a comparison, we discuss the Anderson transition in a normal insulator. With the effective Hamiltonian in the Eq.~(1) of the main text, we use a new set of parameters to describe a normal insulator. The geometry of the sample shown in \cref{Figs1}(a) is the same as Fig.~1(a) in the main text.

Then, we investigate the Hall conductance of the sample to further identity it is a normal insulator. Compared with the axion insulator in Fig.~3, the normal insulator manifests itself in the zero Hall conductance contributed from each layer in \cref{Figs1}(b). As we have done in the main text, we consider a 3D long bar sample of length $L_y$ and widths $L_{x}=L_{z}=L$ with periodic boundary condition in the $x$ direction and open boundary condition in the $z$ direction to calculate the localization length. The result in \cref{Figs1}(c) only shows a critical point, which indicates a 3D Anderson metal-insulator transition, and the 2D critical phase in the axion insulator disappears here. Therefore, we propose to probe the axion insulator state by investigating the universal signature of 2D quantum-Hall-type critical behaviors in the 3D magnetic TI. 
\section{Discussions in an antiferromagneic topological insulator $\mathbf{MnBi_2Ti_4}$}
\begin{figure}[htbp]
    \includegraphics[scale=0.3]{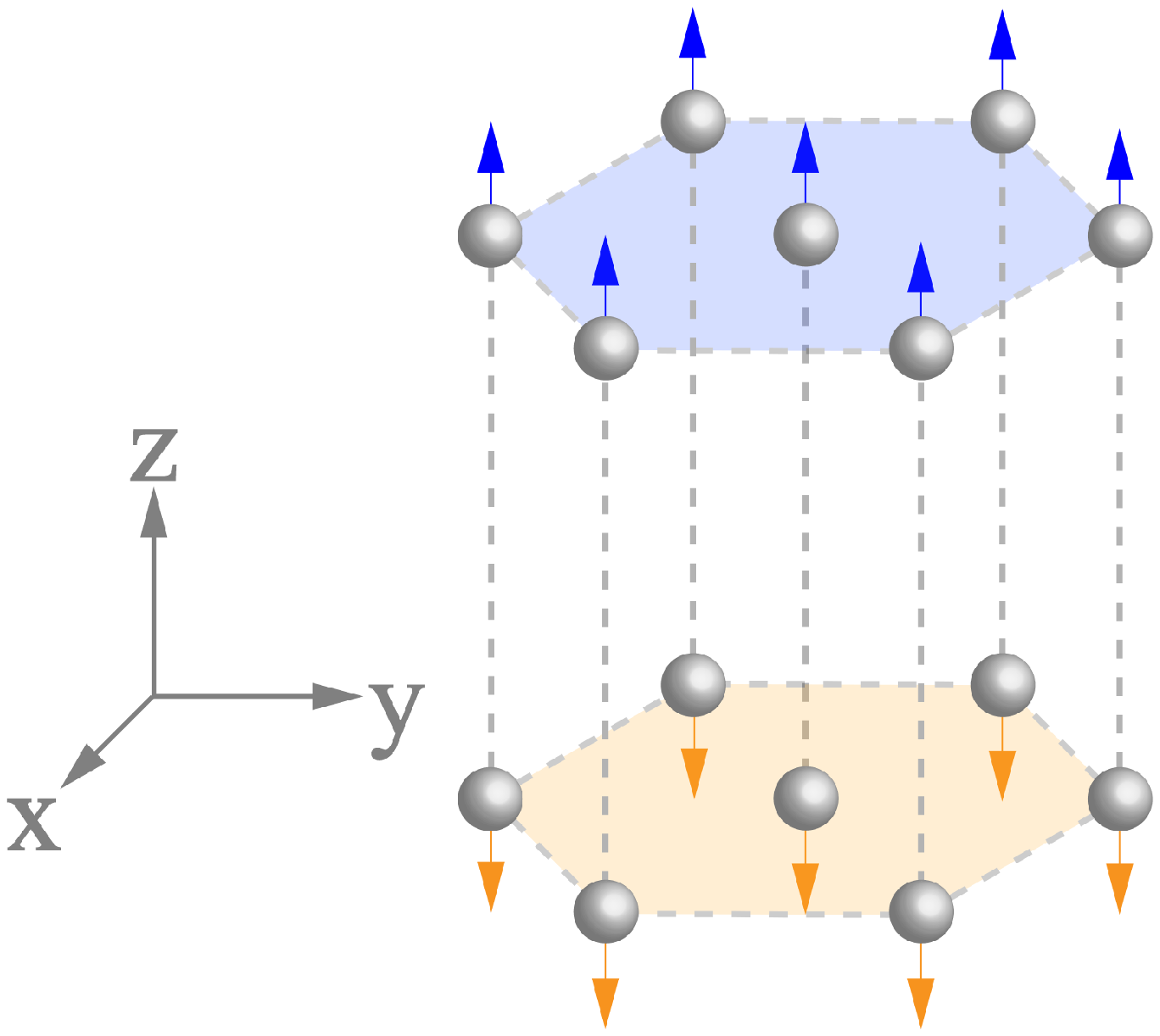}
    \caption{(color online) Lattice structure of the antiferromagneic topological insulator $\mathrm{MnBi_2Ti_4}$ between two adjacent layers.
    }
    \label{mnbite_device}
  \end{figure}
In the main text, we have investigated the disorder-induced Anderson transition of an axion insulator consisting of a 3D time-reversal invariant topological insulator with antiparallel magnetization alignment surfaces, and found a 2D phase transition between the axion insulating phase and the Anderson insulating phase, which does not occur in normal insulators. It is necessary to repeat the finite-size scaling in an antiferromagneic (AFM) topological insulator $\mathrm{MnBi_2Ti_4}$ to confirm its model-independent property. The AFM $\mathrm{MnBi_2Ti_4}$ is fabricated by stacking septuple layers with opposite magnetization between the neighbouring septuple layers [See \cref{mnbite_device}]. Besides, it forms an axion insualtor (quantum anomalous Hall insulator) for even (odd) septuple layers~\cite{MnBiTe_FPC_wangjing}. The effective low-energy model for a AFM $\mathrm{MnBi_2Ti_4}$ reads~\cite{MnBiTe_FPC_wangjing,dassarma},
\begin{equation}
    H=\sum_{i=1}^{4} d_{i}(\mathbf{k}) \Gamma_{i}+ M(z) \cdot s_z\otimes \sigma_0\label{mnbite_H}
\end{equation}
where $d_{1}=A_{1}k_{x}$, $d_{2}=A_{1}k_{y}$, $d_{3}=A_{2}k_{z}$, and $d_{4}=M_0-B_{1} k_{z}^{2}-B_{2}\left(k_{x}^{2}+k_{y}^{2}\right)$. $\Gamma_{i}=s_{i} \otimes \sigma_{1}$ for $i=1,2,3$, and $\Gamma_{4}=s_{0} \otimes \sigma_{3}$. $s_{i}$ and $\sigma_{i}$ are the Pauli matrices for the spin and orbital degrees of freedom. The last term of \cref{mnbite_H} characterize the magnetization of the $z$-th layer by $M(z)=\left[-2\cdot mod(z,2)+1\right]\cdot M_0$, where $mod(...)$ returns the rest of a division.

\begin{figure}[htbp]
    \includegraphics[scale=0.3]{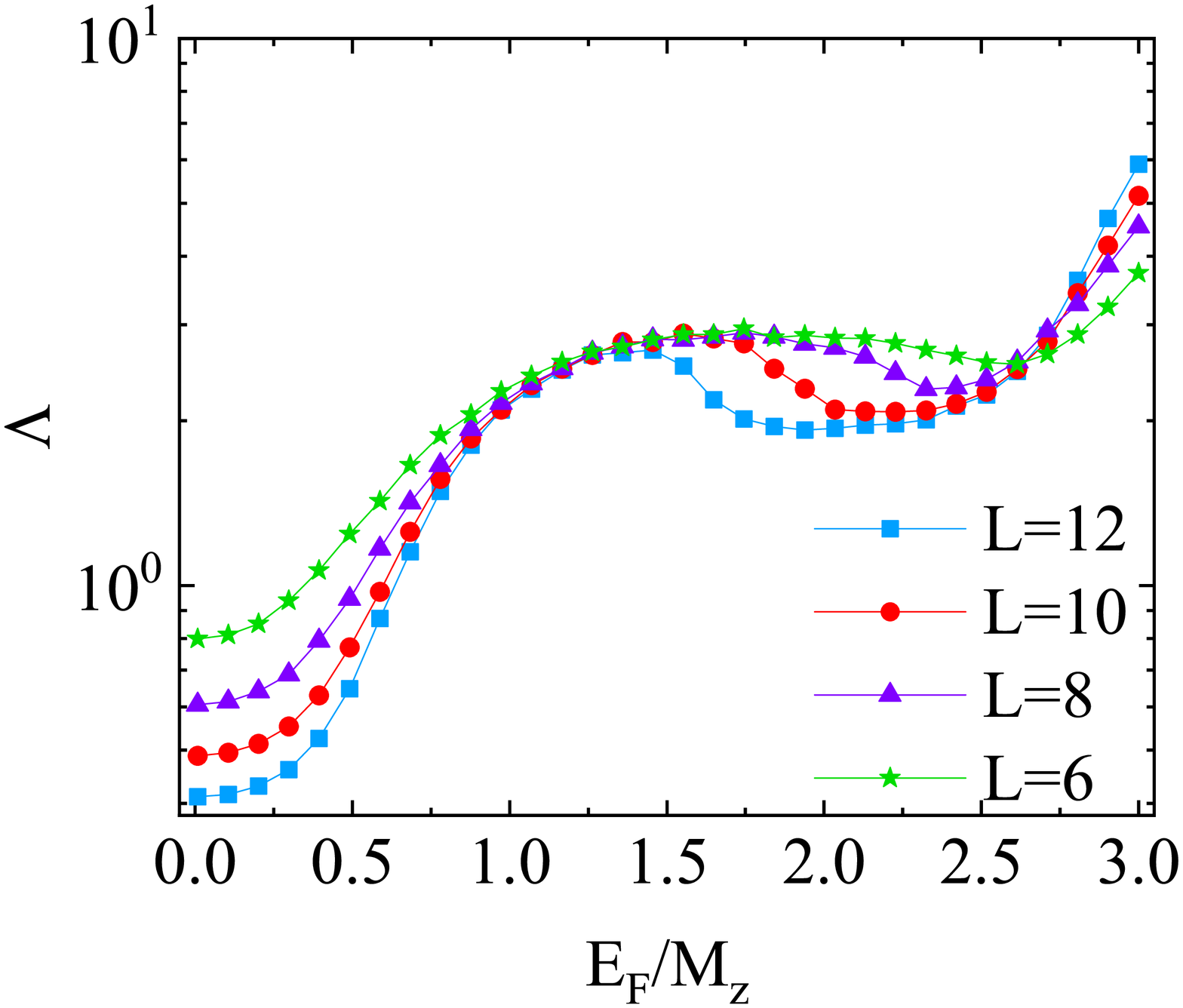}
    \caption{(color online) Renormalized localization length $\Lambda=\lambda(L)/L$ against the Fermi energy $E_{F}/M_{z}$ with the disorder strength $W=1.5$. The curves corresponds to different sample widths $L$. Here, the $x$ direction is periodic and the $z$ direction is open. Parameters of the Hamiltonian are $A_{1}=A_{2}=0.55$, $B_{1}=B_{2}=0.25$, $M_0=0.3$, and $M_{z}=0.12$.
    }
    \label{mnbite_lambda}
  \end{figure}

Based on even-layer $\mathrm{MnBi_2Ti_4}$, we calculate the renormalized localization length as we have done in the main text, and show it in \cref{mnbite_lambda}. Compared with Fig.~2(a) in the main text, \cref{mnbite_lambda} shows the axion insulator $\mathrm{MnBi_2Ti_4}$ undergoes multiple phase transitions with increasing Fermi energy $E_F$ similarly. To be specific, for $W=1.5$ in \cref{mnbite_lambda}, one can identify an axion insulator phase with $d\Lambda/dL<0$ when $|E_F/M_z|\lesssim 1$, where the Fermi energy $E_F$ is within the the surface Dirac gap $M_z$. With increasing the Fermi energy, 
we also find that the system goes through a critical phase where $d\Lambda/dL=0$ and arrive at a normal insulator phase, just like the model we use in the main text. Therefore, we conclude that the 2D phase transition of an axion insulator is model-independent.
